\documentclass[10pt,journal,compsoc]{IEEEtran}
\ifCLASSOPTIONcompsoc
  \usepackage[nocompress]{cite}
\else
  \usepackage{cite}
\fi

\ifCLASSINFOpdf
   \usepackage[pdftex]{graphicx}
\else
\fi
\usepackage{bm}

\usepackage{amssymb}
\usepackage{amsfonts}
\usepackage[margin=2.5cm]{geometry}
\usepackage{graphicx}
\usepackage[ruled, linesnumbered]{algorithm2e}
\usepackage{multirow}
\usepackage{url}
\usepackage[table]{xcolor}
\usepackage{amsmath}
\graphicspath{{./}{images/}}

\usepackage{array}  
\newcolumntype{H}{>{\setbox0=\hbox\bgroup}c<{\egroup}@{}} 

\usepackage{algorithmic}

\begin{document}

\title{From Spectra to Localized Networks: A Reverse Engineering Approach}
\author{Priodyuti~Pradhan 
        and~Sarika~Jalan
\IEEEcompsocitemizethanks{\IEEEcompsocthanksitem Priodyuti~Pradhan was with the Complex Systems Lab, Discipline of 
Physics, Indian Institute of Technology Indore, Khandwa Road, Simrol, Indore-453552, India. 
E-mail: priodyutipradhan@gmail.com
\IEEEcompsocthanksitem Sarika Jalan is with the Complex Systems Lab, Discipline of 
Physics, Indian Institute of Technology Indore, Khandwa Road, Simrol, Indore-453552, India.
E-mail: sarikajalan9@gmail.com
}
}

\markboth{Journal of \LaTeX\ Class Files,~Vol.~14, No.~8, March~2020}%
{Shell \MakeLowercase{\textit{et al.}}: Bare Demo of IEEEtran.cls for Computer Society Journals}

\IEEEtitleabstractindextext{%
\begin{abstract}
Understanding the localization properties of eigenvectors of complex networks is important to get insight into various structural and dynamical properties of the corresponding systems. Here, we analytically develop a scheme to construct a highly localized network for a given set of networks parameters that is the number of nodes and the number of interactions. We find that the localization behavior of the principal eigenvector (PEV) of such a network is sensitive against a single edge rewiring. We find evidences for eigenvalue crossing phenomena as a consequence of the single edge rewiring, in turn providing an origin to the sensitive behavior of the PEV localization. These insights were then used to analytically construct the highly localized network for a given set of networks parameters. The analysis provides fundamental insight into relationships between the structural and the spectral properties of networks for PEV localized networks. Further, we substantiate the existence of the eigenvalue crossing phenomenon by considering a linear-dynamical process, namely the ribonucleic acid (RNA) neutral network population dynamical model. The analysis presented here on model networks aids in understanding the steady-state behavior of a broad range of linear-dynamical processes, from epidemic spreading to biochemical dynamics associated with the adjacency matrices.
\end{abstract}

}

\maketitle

\IEEEdisplaynontitleabstractindextext

\IEEEpeerreviewmaketitle

\IEEEraisesectionheading{\section{Introduction}\label{sec:introduction}}

\IEEEPARstart{N}etworks are composed of interconnected units that interact with each other forming the underlying infrastructures for different dynamical systems \cite{graph_architecture_2018}. The exact pattern of interconnection between these units can take on various forms that dictate the functionality of the corresponding system \cite{rev_Strogatz_2001}. The relationship between the interconnected architecture and functionality is essential to understand the questions pertaining to spreading processes in various real-world dynamical systems, for instance, how the virus spreads nationwide, how information spreads through the social networks or how neurons interact to perform specific functions over the brain networks \cite{rev_dynamical_process_2012, graph_architecture_2018}. An important microscopic question concerns whether a few units of a system participate significantly, and the rest of the others have a tiny contribution, or all the units have the same amount of contribution to a dynamical process \cite{neurons_localization_2014, Goltsev_prl2012}. For instance, during a disease spread, it is important to investigate whether a portion of the network is affected more than the other parts. Whether a small perturbation remains restricted to the vicinity of the source unit or reaches to the remote units, and which properties of a network, for example brain network, allow different regions to process information over different timescales \cite{Goltsev_prl2012, Allesina_nat2015, universality_dynamics_2013, neurons_localization_2014}. Spectral (eigenvalues and eigenvectors) properties of the interaction matrices have been shown to be useful by providing important clues on the interplay between interconnection architecture and network dynamics \cite{pevec_nat_phys_2013, spectrum_controlling_2015, spectra_structure_2000, spectrum_review_2018}. 

A promising paradigm that assists in understanding the spreading phenomena is the localization behavior of eigenvectors of matrices associated with the networks \cite{Goltsev_prl2012, Allesina_nat2015, neurons_localization_2014,brain_networks2013, hierarchical_2017, disease_localization_2018, neural_networks_2015, mieghem_loc_epidemic_2018, biological_network_loc_2017, community_2015, dima_google2015, localization_in_mat_2016}. Localization of an eigenvector refers to a state when a few components of the vector take very high values while the rest of the components take small values independent of the network size. Specifically, the eigenvector corresponding to the largest eigenvalue referred to as principal eigenvector (PEV) of the adjacency matrix approximates the steady-state behavior of the linear-dynamical process on the network ranging from epidemic spreading to biochemical dynamics \cite{pevec_nat_phys_2013}. The epidemic spreading phenomenon is the basis for a broad class of dynamical processes, and a cornerstone feature of epidemic processes is the presence of the so-called epidemic threshold \cite{rev_dynamical_process_2012}. Below the threshold, the disease does not spreads, and above the threshold, the disease spreads across the population. This threshold is inversely proportional to the largest eigenvalue of the adjacency matrix \cite{rev_dynamical_process_2012, epidemic_threshold2003}. In other words, the largest eigenvalue provides a threshold for such spreading processes. However, it does not provide insights into the behavior of the spreading process in the steady-state, i.e., how the nodes of a network are affected during the disease spread. On characterizing the network properties that enhance the localization of PEV, it is found that a few interconnected units participate significantly with the rest of the units contributing very less in the steady-state \cite{Goltsev_prl2012, brain_networks2013, hierarchical_2017}. Hence, the network properties which enhance PEV localization can implicitly restrict the linear-dynamics in a smaller section of the network in the steady-state. A fundamental question at the core of the structural-dynamical relation is that: How can one construct a graph structure to affect the outcome of a linear-dynamical process in a desired manner? Construction of a localized network structure can help us to engineer the system's architecture, enhance robustness, as well as can provide insights into the underlying mechanism of evolution of structural patterns of real-world complex systems \cite{brain_networks2013, sensitivity_dynamics_2017, eigen_optimization_2016}. By understanding the relationship between spreading processes on networks and structural properties, one can artificially construct a network structure with control over linear-dynamical processes. Therefore, it is important to understand the relationships between structural properties of network and PEV localization of the underlying adjacency matrices.

Taking a clue from the relationship between structural and spectral properties of numerically achieved localized network structure \cite{evec_localization_2017}, the present study provides an analytical method for the construction of a highly localized network structure for a given set of network parameters. We show that the highly localized network structure is accompanied by sensitivity in the localization behavior of PEV against a single edge rewiring. Moreover, we find evidences for eigenvalue crossing phenomena as a consequence of single edge rewiring. This finding, in turn, provides an origin to the sensitivity in the localization behavior of PEV against a single edge rewiring. The investigation can be summarized as follows: First, we demonstrate that the highly localized network structure for a given set of network parameters can be constructed by combining a wheel and a random regular graph. For a given set of networks parameter, that is the number of nodes and the number of connections; we analytically obtain the size of the wheel and the random regular graph to construct the PEV localized networks. Second, we demonstrate that such PEV localized networks show the eigenvalue crossing phenomenon.  Thereafter, we establish a relationship between this eigenvalue crossing phenomenon and the sensitive behavior of the PEV localization. Third, we substantiate the eigenvalue crossing phenomenon by using the RNA neutral network population dynamical model. 

The article is designed as follows: Section \ref{Theoretical Framework} describes the notations and definitions of the mathematical terms and section \ref{Motivation} provides the motivation of the developed method. Section \ref{Results} illustrates the analytical method for the construction of a PEV localized network structure. Subsection \ref{RNA} describe the results for the steady-state behavior of the RNA neutral network population dynamical model on localized networks. Finally, section \ref{conclusion} summarizes our work and discusses various open problems for further investigations.

\section{Preliminaries} \label{Theoretical Framework}
We represent a finite graph, $\mathcal{G} =\{V,E\}$, where $V=\{v_1, v_2,\ldots,v_n\}$ is the set of vertices (nodes) and $E=\{e_1, e_2,\ldots,e_{m}|e_{p}=(v_i, v_j), p=1,2,\ldots,m\} \subseteq U$ is the set of edges (connections). We define the universal set $U = V \times V=\{(v_i, v_j)| v_i, v_j \in V\; \mbox{and}\; i \neq j\}$ which contains all possible unordered pairs of vertices excluding the self-loops. The complementary set of the edges can be defined as $E^c = U - E=\{(v_i, v_j)| (v_i, v_j) \in U\; \mbox{and}\; (v_i, v_j) \notin E\}$ i.e., $E\cap E^c=\varnothing $ and $E \cup E^c= U$. We denote the adjacency matrix corresponding to $\mathcal{G}$ as ${\bf A} \in \mathbb{R}^{n \times n}$ which can be defined as 
\begin{equation}\nonumber
a_{ij}=
   \begin{cases}
     1       & \quad \text{if nodes $i$ and $j$ are connected} \\
     0  & \quad \text{ Otherwise}\\
   \end{cases}
\nonumber   
\end{equation}
The $|V|=n$ and $|E|=m$ represent the number of nodes and number of edges in $\mathcal{G}$, respectively, and thus $|E^c|=\frac{n(n-1)}{2}-m$. The number of edges to a particular node is referred as its degree denoted as $d_i = \sum_{j=1}^{n} a_{ij}$. The average degree of the network is denoted by $\langle k \rangle$ = $\frac{1}{n}\sum _{i=1}^{n} d_{i}$. We refer the maximum degree node or the hub node of $\mathcal{G}$ as $k_{max} = \max_{1\leq i \leq n} d_i$. Here, ${\bf A}$ is a real symmetric matrix, hence, it has a set of orthonormal eigenvectors $\{\bm{x}_1, \bm{x}_2, \cdots, \bm{x}_n\}$ corresponding to the real eigenvalues $\{\lambda_1, \lambda_2, \ldots, \lambda_n\}$ such that $\lambda_1> \lambda_2\geq \ldots\geq \lambda_n$. The eigenvector ($\bm{x}_1$) corresponding to $\lambda_1$ is referred as the principal eigenvector (PEV) \cite{eigenspaces_of_graphs_1997}.
Moreover, the edge weights of ${\bf A}$ are non-negative ($a_{ij} \geq 0$), and in our current study network is always connected. Thus, ${\bf A}$ is a non-negative and irreducible matrix  \cite{matrix_analysis}. Hence, we know from the Perron-Frobenius theorem that all the entries in PEV of ${\bf A}$ are positive, and $\lambda_1$ is simple (non-degenerate) \cite{matrix_analysis}. 

We use the inverse participation ratio (IPR) to quantify the localization as well as delocalization behavior of eigenvectors in complex networks \cite{castellano_localization_2017, Goltsev_prl2012, brain_networks2013, hierarchical_2017}. This measure had been introduced to quantify participation of atoms in normal mode and is similar to the fourth moment in statistics \cite{ipr_11_1972,ipr_1_1980, ipr_2_1970}. We calculate the IPR value ($Y_{\bm{x}_{j}}$) of an orthonormal eigenvector ($\bm{x}_{j}=((x_j)_{1},(x_j)_{2},\ldots,(x_j)_{l},\ldots,(x_j)_{n})^{T}$) of ${\bf A}$ as follows:
\begin{equation} \label{eq_IPR}
Y_{\bm{x}_{j}}=\sum_{l=1}^n (x_{j})_{l}^4 
\end{equation}
where $(x_{j})_{l}$ is the $l^{th}$ component of $\bm{x}_j$. A most localized eigenvector ($\bm{x}_{j}=(1,0,\ldots,0)$) yields an IPR value equal to $Y_{\bm{x}_j} = 1$, whereas the delocalized eigenvector ($\bm{x}_{j}=(\frac{1}{\sqrt{n}},\frac{1}{\sqrt{n}},\ldots,\frac{1}{\sqrt{n}})$) has $Y_{\bm{x}_j}=\frac{1}{n}$. In general, for a network, eigenvector is said to be localized if $Y_{\bm{x}_j}=\mathcal{O}(1)$ and delocalized if $Y_{\bm{x}_j} \rightarrow 0$ as $n\rightarrow \infty$ \cite{Goltsev_prl2012,ipr_11_1972}.

\section{Motivation} \label{Motivation}
For a graph with each node only having self-loop without having any interaction with any other node, the corresponding adjacency matrix will be an identity matrix and for which we can choose $\bm{x}_{1}=(1,0,\ldots,0)$ yielding $Y_{\bm{x}_{1}} = 1$. However, for connected networks, all entries of the PEV should be positive (from the Perron-Frobenius theorem). Hence, IPR of the PEV should be less than $1$ for $n \geq 2$. Next, if we consider a star graph having $n$ nodes (and thus $n-1$ connections), $\bm{x}_{1}^{\mathcal{S}}=\biggl(\frac{1}{\sqrt{2}},\frac{1}{\sqrt{2(n-1)}},\ldots, \frac{1}{\sqrt{2(n-1)}}\biggr)$ and hence, $Y_{\bm{x}_{1}^{\mathcal{S}}} = \frac{1}{4} + \frac{1}{4(n-1)}$. Considering a wheel graph with $n$ nodes having $2(n-1)$ connections \cite{eigval_wheel_graph}, we have $\bm{x}_{1}^{\mathcal{W}}=(\frac{1}{\beta},\frac{\alpha}{\beta},\ldots,\frac{\alpha}{\beta})$ where $\alpha=\sqrt{n}+1/(n-1)$, $\beta=\sqrt{1+(\sqrt{n}+1)^2/(n-1)}$, and $Y_{\bm{x}_{1}^{\mathcal{W}}} = \frac{1}{4}\frac{(n-1)^2}{(n+\sqrt{n})^2} + \frac{(\sqrt{n}+1)^4}{4(n-1)(n+\sqrt{n})^2}$. Hence, for $n \rightarrow \infty$, we get $Y_{\bm{x}_{1}^{\mathcal{S}}} = Y_{\bm{x}_{1}^{\mathcal{W}}} \approx 0.25$, and PEV is localized for both the star and the wheel networks. However, for a regular network (with each node having the same degree) of $n$ nodes and each node having degree $\kappa$ yielding $m=\frac{n\kappa}{2}$, we have $\bm{x}_{1}=(\frac{1}{\sqrt{n}},\frac{1}{\sqrt{n}},\ldots,\frac{1}{\sqrt{n}})$ (Theorem 6 \cite{miegham_book2011}) yielding, $Y_{\bm{x}_1}=\frac{1}{n}$. Further, for other model networks such as Erd\"os-R\'enyi (ER) random or scale-free (SF) networks ($n$ number of nodes, $m$ number of edges and $m>>n$) \cite{network_sec_2016}, it is difficult to find a closed functional form of PEV and thereby it is hard to find the IPR value analytically. It has been reported that for ER random networks with each node having the same expected degree, we get a delocalized PEV \cite{deloc_pev}. In contrast, for SF networks, the presence of hub nodes and power-law degree distribution lead some amount of localization in the PEV, and IPR value while being larger than that of the ER random networks is much lesser than that of the star networks \cite{Goltsev_prl2012, satorras_localization2016, castellano_localization_2017}. Although for the star, wheel, regular, ER random, and scalefree networks with the same $n$ have different IPR values. In other words, for a given number of nodes and connections, some network structures have localized PEV, whereas some other network structures lead to delocalized PEV. An important observation is that although all the networks have the same number of nodes, have a different number of edges and diverse connection pattern. Here, we ask a general question, for a given $n$ and $m$ (or $\langle k \rangle$) values, how can we construct a network structure having a highly localized PEV?

From Ref. \cite{evec_localization_2017}, we know that there exists a network structure that is different from the star network, having a highly localized PEV, which we can construct by using an optimization process. However, for a large size network of $n$ number of nodes and $m$ number of edges, the optimized edge rewiring process of Ref. \cite{evec_localization_2017} is computationally intractable to construct a highly localized network structure. The current article focuses on developing an analytical approach for the construction of a highly localized network structure from given values of $n$ and $m$ and avoids the optimized edge rewiring process.

\section{Results}\label{Results}
\begin{table}[b]
\centering
\begin{tabular}{|c|c|c|c|HHc|>{\bfseries}c|HHc|c|}
\hline
\multirow{2}{*}{}No. &$\mathcal{G}$&$n_1$ & $n_2$ & $\langle k\rangle^{\mathcal{C}_1}$ & $\langle k\rangle^{\mathcal{C}_2}$
&$k_{max}$& $Y_{\bm{ x}_1}^{\mathcal{G}}$& $\lambda_1^{\mathcal{G}}$&$\lambda_2^{\mathcal{G}}$&$\lambda_1^{\mathcal{C}_1}$ &$\lambda_1^{\mathcal{C}_2}$ \\  
\hline
1.&SF-ER&500&500 &10 &4 &69 &0.003& 11.03&10.09& 10.09&11.03\\
2.&$\mathcal{W}-\mathcal{R}$ &24 &500 & & &23&0.002&6&5.9&5.89&6\\
\hline
3.&SF-ER& 500 &500&8&4&68&0.08&10.27&9.59&10.24&9.58\\
4.&$\mathcal{W}-\mathcal{R}$ &26 & 500& & &25&0.17&6.1&5.99&6.09&6\\
\hline
\end{tabular}
\caption{Structural and spectral properties of two components ($\mathcal{C}_1$ and $\mathcal{C}_2$) separately, and the one achieved by connecting them through a link. We consider ER random graph, scalefree (SF), wheel ($\mathcal{W}$) and random regular ($\mathcal{R}$) networks as individual component. Satisfying $\lambda_1^{\mathcal{C}_1}>\lambda_1^{\mathcal{C}_2}$ leads to a localized PEV and for  $\lambda_1^{\mathcal{C}_1}<\lambda_1^{\mathcal{C}_2}$ yields delocalized PEV of the combined graph.}
\label{graph_construction}
\end{table}
An earlier work have shown that the most localized network structure should consist of two subgraphs ($\widetilde{\mathcal{C}}_1$ and $\widetilde{\mathcal{C}}_2$) with one having a hub node \cite{evec_localization_2017}. Here, we show that combining any two subgraphs, say $\mathcal{C}_1$ and $\mathcal{C}_2$, with one of them containing a hub node may not always produce a localized PEV for the combined network (Table \ref{graph_construction}). A rather interesting observation is that the combined network ($\mathcal{G}$) having localized PEV obeys the following eigenvalue relationship ($\lambda_1^{\mathcal{C}_1} > \lambda_1^{\mathcal{C}_2}$) between its individual sub-graphs or components (Table \ref{graph_construction}). In the following, we provide an analytical framework revolving around the existence of the eigenvalue relationship between the individual component to generate a highly localized network structure for a given set of network parameters.

\subsection{Analytical method for construction of localized network by combining wheel and random regular graphs}\label{Analytical}
First of all, we have made a partition to the given values of $n$ and $m$ into two groups ($\mathcal{C}_1$ and $\mathcal{C}_2$) such that
\begin{eqnarray}
n = n_1 + n_2 +1 \label{nodes_relation}\\
m = m_1 + m_2 +2 \label{edge_relation}\\
\lambda_1^{\mathcal{C}_1} > \lambda_1^{\mathcal{C}_2} \label{feval_relation}
\end{eqnarray}
where ($n_1$,$m_1$) and ($n_2$,$m_2$) are unknown values corresponding to the size of the two graph components ($\mathcal{C}_1$ and $\mathcal{C}_2$), respectively such that these sub graphs satisfy Eqs. (\ref{nodes_relation}), (\ref{edge_relation}), and (\ref{feval_relation}). From the numerical simulations in Ref. \cite{evec_localization_2017}, we learn that in the optimized networks, $\widetilde{\mathcal{C}}_1$ contains a hub node while $\widetilde{\mathcal{C}}_2$ has almost a regular structure. We use these clues to choose graph structures of $\mathcal{C}_1$ and $\mathcal{C}_2$. The closest structures corresponding to $\widetilde{\mathcal{C}}_1$ could be a star, wheel, or friendship graphs \cite{miegham_book2011,eigval_wheel_graph,friendship_graph_2013}. Whereas, for the $\mathcal{C}_2$ component, we choose a random regular structure \cite{random_regular_graph}. Note that the $\widetilde{\mathcal{C}}_1$ and $\widetilde{\mathcal{C}}_2$ components of the optimized structure is not exactly the same as a wheel and a random regular structure.

\begin{figure}[t]
\begin{center}
\includegraphics[width=3.2in, height=0.75in]{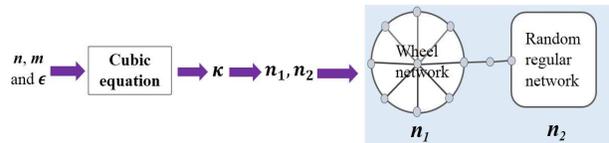}
\caption{Method to construct the PEV localized network through the solution of cubic equation. Given the input parameters (number of nodes ($n$), connections ($m$) and $\epsilon<1$), the coefficients of the cubic equation provide the roots (average degree $\kappa$) for the calculation of the size of wheel ($\mathcal{W}$) and random regular ($\mathcal{R}$) networks (Eqs. (\ref{vertices_wheel_relation}) and (\ref{vertices_random_relation})). Finally, connecting $\mathcal{W}$ and $\mathcal{R}$ yields the PEV localized network.}  
\label{wheel_graph}
\end{center}
\end{figure}

To construct the PEV localized network ($\mathcal{G}_{new}$) by combining a wheel and a random regular structure by satisfying Eqs. (\ref{nodes_relation}), (\ref{edge_relation}), and (\ref{feval_relation}), we need the information about parameters of the individual component 
($n_1$, $n_2$, $m_1$, and $m_2$) from a given $n$ and $m$ value. Let us denote the wheel graph as $\mathcal{W}=\{V_\mathcal{W},E_\mathcal{W}\}$ which is formed by connecting one node to all the nodes of a cycle graph of size $n_1-1$, where $|V_\mathcal{W}|=n_1$ is the number of nodes and $|E_\mathcal{W}|=m_1=2(n_1-1)$ is the number of edges in $\mathcal{W}$  \cite{eigval_wheel_graph}. Further, let us denote the random regular graph as $\mathcal{R}=\{V_\mathcal{R},E_\mathcal{R}\}$ where $|V_\mathcal{R}|=n_2$ is the number of nodes and $|E_\mathcal{R}|=m_2=\frac{n_2\kappa}{2}$ is the number of edges with each node having degree $\kappa$ such that $\kappa n_2$ is even. We generate the random regular graph using the algorithm in \cite{random_regular_graph}. It is known that for a wheel and random regular graph, the largest eigenvalues are as follows \cite{eigval_wheel_graph,miegham_book2011}
\begin{equation}\label{eig_vals}
\lambda_1^{\mathcal{W}} = 1+\sqrt{n_1}\;\;\mbox{and}\;\; \lambda^\mathcal{R}_{1}=\kappa 
\end{equation}
Let us start with an example of a wheel graph of size $n_1=290$ and hence $m_1=578$ and a random regular graph having $n_2=209$ and each node having the same degree $\kappa=18$, thus $m_2=1881$. From Eq. (\ref{eig_vals}), we know that $\lambda_1^{\mathcal{W}}=18.03$ and $\lambda_1^{\mathcal{R}}=18$, respectively. Next, by combining these two components via a node (Fig. \ref{wheel_graph}), we have $n=500$ nodes (Eq. (\ref{nodes_relation})) and $m=2461$ edges (Eq. (\ref{edge_relation})) and satisfy Eq. (\ref{feval_relation}), hence we know that PEV is localized (\cite{evec_centrality_loc_2020}, Appendix B). Next, we ask the reverse question, for a given ER random network with $n=500$ and $m=2461$ (or $\langle k \rangle=10$), how can we find $n_1$, $n_2$, and $\kappa$ for two individual components such that Eqs. (\ref{nodes_relation}), (\ref{edge_relation}) and (\ref{feval_relation}) hold true? Note that usage of the wheel and the random regular network structures reduces the number of parameters from four ($n_1$, $n_2$, $m_1$, and $m_2$) to three ($n_1$, $n_2$, and $\kappa$) for a given $n$ and $m$. We can approach the problem by choosing a value for $\kappa$, say $12$, and we need an even $n_2\kappa$, hence, let us take $n_2=100$ which gives $m_2 = 600$ and $\lambda_1^{\mathcal{R}}=12$. Therefore, $n_1=399$, $m_1=796$ and $\lambda_1^{\mathcal{W}}=20.97$ and average degree of the combined network comes close to $2.8$ which is far from $10$. Hence, Eqs. (\ref{nodes_relation}) and (\ref{feval_relation}) get satisfied but Eq. (\ref{edge_relation}) does not, and it is tedious to find a solution. The cubic equation in the following is constructed to automate this process and to find a solution.

For the relation $\lambda_1^{\mathcal{W}}>\lambda_1^{\mathcal{R}}$, we consider,
\begin{equation}\label{eval_relation}
\lambda_1^{\mathcal{W}} =\lambda_1^{\mathcal{R}}+\epsilon\;\;\mbox{where} \;\; 0< \epsilon < 1\\
\end{equation}
Here, we use $\epsilon<1$ to demonstrate the eigenvalue crossing phenomenon as a consequence of a single edge rewiring. From Eqs. (\ref{eig_vals}) and (\ref{eval_relation}), we obtain the size of the wheel graph as follows
\begin{equation} \label{vertices_wheel_relation}
n_1=\lceil (\kappa-1+\epsilon)^2\rceil
\end{equation}
where ($\lceil$ $\rceil$) is the ceiling function. Implicitly, Eq. (\ref{vertices_wheel_relation}) tells that for any $\kappa$ ($3 \leq \kappa \leq n_2-2$), if we take $\lceil (\kappa-1+\epsilon)^2\rceil$ as a number of nodes for the wheel graph, the combined graph will satisfy Eq. (\ref{eval_relation}). Importantly, in Eq. (\ref{vertices_wheel_relation}) the number of nodes in the $\mathcal{W}$ component of $\mathcal{G}_{new}$ depends on the average degree of the $\mathcal{R}$ component in $\mathcal{G}_{new}$. Further, from Eqs. (\ref{nodes_relation}) and (\ref{vertices_wheel_relation}) we know that
\begin{equation}\label{vertices_random_relation}
n_2 = n-\lceil(\kappa-1+\epsilon)^2\rceil-1 
\end{equation}
Now, we substitute $m_1=2(n_1-1)$ and $m_2=\frac{n_2\kappa}{2}$ in Eq. (\ref{edge_relation}) and we get, 
\begin{equation}\label{edges_relation}
m =\frac{4n_1+n_2\kappa}{2}\\
\end{equation}
Finally, we rearrange Eq. (\ref{edges_relation}) with the help of Eqs. (\ref{vertices_wheel_relation}) and (\ref{vertices_random_relation}), and arrive to a cubic equation of the form
\begin{equation}\label{cubic_eq}
\kappa^3+b\kappa^2+c\kappa+d =0
\end{equation}
where $b=(-4-2(1-\epsilon))$, $c= ((1-\epsilon)^2+8(1-\epsilon)+1-n)$, and $d= (2m-4(1-\epsilon)^2)$ are the coefficient of the cubic equation in terms of $n$, $m$, and $\epsilon$. Next, roots of the cubic equation can be written from the Cardano's formula \cite{cubic_roots} as follows,
\begin{equation} \label{roots}
\begin{split}
\kappa_1 &= \Delta_1 + \Delta_2 - \frac{b}{3}\\
\kappa_2 &= -\frac{1}{2}(\Delta_1 + \Delta_2)-\frac{i\sqrt{3}}{2}(\Delta_1 - \Delta_2) - \frac{b}{3} \\
\kappa_3 &= -\frac{1}{2}(\Delta_1 + \Delta_2)+\frac{i\sqrt{3}}{2}(\Delta_1 - \Delta_2) - \frac{b}{3}
\end{split}
\end{equation}
such that
\begin{eqnarray}\label{deltas}
\Delta_1 &= \sqrt[3]{- \beta/2 + \sqrt{\Delta}},\;
\Delta_2 = \sqrt[3]{- \beta/2 - \sqrt{\Delta}}
\end{eqnarray}
\begin{figure}[t]
\begin{center}
\includegraphics[width=3in, height=0.8in]{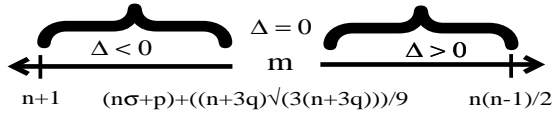}
\caption{Separation of $m$ values based on the behavior of the discriminant value ($\Delta$) of the cubic equation (in Eq. \ref{roots}) for a particular value of $n$. For sparse network $\Delta<0$ and $\Delta \geq 0$ as network becomes dense. Here, $\sigma=(1-\frac{\epsilon}{3})$, $p = \frac{\epsilon^{3}+9\epsilon^{2}+36\epsilon}{27}$, $q=\frac{\epsilon^2+6\epsilon+6}{9}$ and we consider $n \geq 49$.} 
\label{range}
\end{center}
\end{figure}
where $\Delta = \frac{\beta^2}{4}+\frac{\alpha^3}{27}$, $\alpha = \frac{1}{3}(3c-b^2)$, $\beta = \frac{1}{27}(2b^3-9bc+27d)$ and $i=\sqrt{-1}$. Therefore, given a set of $n$ and $m$, we obtain three different possible values for $\kappa$ to partition $n$ and $m$ between two subgraphs while satisfying Eqs. (\ref{nodes_relation}), (\ref{edge_relation}), and (\ref{feval_relation}). There is a possibility to get complex values for $\kappa$. The following analysis presents bounds on the value of $m$ to avoid complex numbers as well as few other unnecessary situations. 

\begin{table*}[t]
\centering
\begin{tabular}{|cc|ccccHHHH|ccccHHHH|}
\hline
\multirow{2}{*}{}$n$ & $m$ & $\kappa_1$ & $n_1$&$n_2$ & $Y_{\bm{ x}_1}$&$\lambda_1^{\mathcal{G}}$&$\lambda_2^{\mathcal{G}}$ &$Y_{\bm{ x}_1}^d$&$m^{\kappa_1}$&$\kappa_2$ &$n_1$&$n_2$ & $Y_{\bm{ x}_1}$&$\lambda_1^{\mathcal{G}}$&$\lambda_2^{\mathcal{G}}$&$Y_{\bm{ x}_1}^d$&$m^{\kappa_2}$\\  
\hline
500&2512 &18&290 & 209&0.22&&&0.005&2461&13 &145 & 354&0.21&&&0.003 &2591\\
520&2630  &19&325 & 194&0.22&&&0.005&2493&13 &145 & 374&0.21&&&0.003 &2721\\
2448&14806&46 &2027 & 420 &0.23&&&0.002&14449&13 &145 & 2302 &0.21&&&0.0004&14993\\
4720&13712&69 &4627 & 92 &0.24&&&0.01&12493 &6 &26 & 4693 &0.17 &&&0.0002&14131\\
10498& 52490&101&10005&492&0.24&&&0.002&45050 & 11&101&10396&0.20&&&0.0001&57380\\
20422& 163376 &138&18775&1646&0.24&&&0.0006&151459 & 17&257&20164&0.22&&&0.00001&171908\\
\hline
\end{tabular}
\caption{Various network parameters and IPR values of PEV for a given set of $n$ and $m$. From Eq. (\ref{roots}), we decide $\kappa$, $n_1$ and $n_2$. Thereupon, we construct a wheel graph of size $n_1$ and a random regular graph of size $n_2$, and join them with a node. This method leads to a highly localized PEV. We consider here $\epsilon=0.02$.}
\label{analytic-verifications}
\end{table*}
We know that the discriminant ($\Delta$) leads to a change in the nature of the roots. One can notice from Eq. (\ref{discriment_analysis}) that $\Delta$ is a function of $n$ and $m$. Furthermore, we know that for a given value of $n$, the value of $m$ can vary between $n+1$ to $n(n-1)/2$. Hence, by varying $m$, we get $\Delta$ as a function of $n$. It turns out that as $m$ varies {\bf ($n+1 \leq m \leq n(n-1)/2$)} for a given $n$ value, the nature of the roots changes yielding real or complex values for $\kappa_i$'s. However, we do not know the exact relation between $m$ and $\Delta$.
It is known that (a) $\Delta=0$ yields three real roots in which at least two are equal, (b) $\Delta>0$ gives one real root and other two complex conjugate roots, (c) $\Delta<0$ yields three unequal real roots \cite{cubic_roots}. To know the behavior of the discriminant as $m$ changes for a particular value of $n$, we analyze $\Delta$ in Eq. (\ref{deltas}) of the cubic equation as; 
\begin{equation}\label{discriment_analysis}
\Delta = (m-n\sigma-p)^2 -\biggl(\frac{n}{3}+q\biggr)^{3}
\end{equation} 
for $m=n+1,n+2,\ldots,\frac{n(n-1)}{2}$ and where $\sigma=(1-\frac{\epsilon}{3})$, $p = \frac{\epsilon^{3}+9\epsilon^{2}+36\epsilon}{27}$, $q=\frac{\epsilon^2+6\epsilon+6}{9}$ and we consider $n \geq 49$ (Appendix). Analyzing the discriminant reveals that for 
\begin{equation}\label{range_for_edges}
m= (n\sigma+p) + \frac{(n+3q)\sqrt{3(n+3q)}}{9} 
\end{equation}
(a) $\Delta=0$ (Appendix). Further, from Eq. (\ref{range_for_edges}), we find the lower and upper bounds of $m$ for which $\Delta<0$ and $\Delta>0$ as follows
\begin{eqnarray} 
n+1 \leq m \leq \biggl \lceil (n\sigma+p-1)+ \frac{(n+3q)\sqrt{3(n+3q)}}{9} \biggr \rceil \nonumber 
\end{eqnarray}
\begin{eqnarray}
\biggl \lceil (n\sigma+p+1)+ \frac{(n+3q)\sqrt{3(n+3q)}}{9} \biggr \rceil \leq m \leq \frac{n(n-1)}{2}  \nonumber
\end{eqnarray}
The ranges of $m$ illustrates that as network becomes dense, $\Delta$ becomes greater or equal to zero (Fig. \ref{range}). From Eqs. (\ref{discriment_analysis}) and (\ref{range_for_edges}), one can see that $\Delta=0$ appears when $m$ is a real with fractional part. 
However, in our case, $m$ represents the number of edges in $\mathcal{G}_{new}$ and is a positive integer. Hence, $\Delta=0$ can never appear. Further analysis of the discriminant reveals that for (b) $\Delta > 0$, $n_1$ calculated from $\kappa_1$ (in Eq. (\ref{vertices_wheel_relation})) is always larger than the given value of $n$. Hence, we can not use $\kappa_1$ to find $n_1$ and $n_2$ in Eqs. (\ref{vertices_wheel_relation}) and (\ref{vertices_random_relation}) for the construction of $\mathcal{G}_{new}$ (Appendix). Finally, we investigate the case (c), which corresponds to three unequal real roots in Eq. (\ref{roots}) (Appendix). However, among the three roots, we can use two different roots to divide the number of nodes in two different groups such that the entire network has a localized PEV. The first way is to consider a sparse regular structure with the smaller size for the wheel graph, and the second way is to consider a dense regular structure with the larger size for the wheel graph. Hence, the coefficients of the cubic equation for a given input parameters $n$ and $m$ give us the subgraph parameters analytically for the construction of PEV localized networks (Fig. \ref{wheel_graph}).

\begin{algorithm}[!htb]
\label{algo}
\SetCommentSty{sf}
\caption{loc\_PEV($n$, $m$, $\epsilon$)}
\Indp 
Evaluate the roots of the cubic equation 
\begin{equation}\nonumber
\begin{split}
\mathfrak{\kappa}_1 &= \Delta_1 + \Delta_2 - \frac{b}{3} \\
\mathfrak{\kappa}_2 &= -\frac{1}{2}(\Delta_1 + \Delta_2)-\frac{i\sqrt{3}}{2}(\Delta_1 - \Delta_2) - \frac{b}{3}\\
\mathfrak{\kappa}_3 &= -\frac{1}{2}(\Delta_1 + \Delta_2)+\frac{i\sqrt{3}}{2}(\Delta_1 - \Delta_2) - \frac{b}{3}
\end{split}
\end{equation}
where
\begin{equation}\nonumber
\begin{split}
\Delta_1 &= \sqrt[3]{- \beta/2 + \sqrt{\Delta}},\;
\Delta_2 = \sqrt[3]{- \beta/2 - \sqrt{\Delta}}
\end{split}
\end{equation}
such that $\Delta=\frac{\beta^2}{4} + \frac{\alpha^3}{27}$, $\alpha = \frac{1}{3}(3c-b^2)$, $\beta = \frac{1}{27}(2b^3-9bc+27d)$, $i=\sqrt{-1}$ where $b=(-4-2(1-\epsilon))$, $c= ((1-\epsilon)^2+8(1-\epsilon)+1-n)$ and $d= (2m-4(1-\epsilon)^2)$\\
For each $\kappa_i$ we can calculate
\begin{equation}\nonumber 
\begin{split}
n_1 &=\lceil (\kappa-1 + \epsilon)^2\rceil \\
n_2 &= n- n_1 - 1 
\end{split}
\end{equation}
and generate a wheel graph with size $n_1$ and a random regular network of size $n_2$ having degree $\kappa_i$ and combined via a node. 
\Indm
\end{algorithm}
Table \ref{analytic-verifications} verifies the theoretical method of arranging the graph components into two different ways while satisfying Eqs. (\ref{nodes_relation}), (\ref{edge_relation}), and (\ref{feval_relation}). For a given set of values for $n$ and $m$, we calculate the average degree of regular graph ($\kappa_{1}$) from the Eq. (\ref{roots}). From Eqs. (\ref{vertices_wheel_relation}) and (\ref{vertices_random_relation}), we calculate $n_1$ and $n_2$ values which in turn provide us the size of the wheel and the random regular graphs, respectively, while satisfying Eq. (\ref{eval_relation}). The combined graph has a highly localized $\bm{x}_{1}$. Similarly, the root $\kappa_2$ can be calculated by the same procedure (Algorithm \ref{algo}). The value of $Y_{\bm{ x}_{1}}$ obtained from the analysis is close to that obtained from the optimized edge rewiring process \cite{evec_localization_2017}. 

\begin{figure}[t]
\begin{center}
\includegraphics[width=3in, height=1.8in]{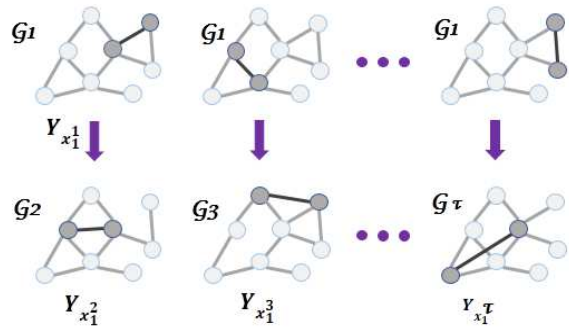}
\caption{$\mathcal{G}_1$ is the initial wheel random regular network. For the random edge rewiring, we remove one edge uniformly at random from $\mathcal{G}_1$ and add it randomly. The new network is represented as $\mathcal{G}_2$. We record the IPR value ($Y_{\bm{x}_1^{2}}$) and restore the edge back to its original position. This two step rewiring process is repeated for $\tau$ times and the sequence of IPR values as $\{Y_{\bm{x^{1}}_1},Y_{\bm{x^{2}}_1},\ldots,Y_{\bm{x^{\tau}}_1}\}$ is recorded.}
\label{schematic_random_edge_rewiring}
\end{center}
\end{figure} 
The analysis presented in this section demonstrates that by considering $\lambda_{1}^{\mathcal{W}} > \lambda_1^{\mathcal{R}}$, one can produce a network structure having a highly localized PEV for a given set of $n$, $m$, without using an optimized edge rewiring process. Note that from the above, we get the root ($\kappa$) of the cubic equation as the average degree of the random regular graph, which may be a real value. However, we round it up to the nearest integer such that $\kappa n_2$ is even. Further, during the calculation of the size of the wheel graph, we take the ceiling function ($\lceil \; \rceil$) in Eq. (\ref{vertices_wheel_relation}). Hence, there may exist changes in the number of edges of the combined graph from the given number of edges. 

\subsection{Analysis of the eigenvectors angles: signature of eigenvalue crossing}\label{eval_crossing}
To capture the impact of single edge rewiring on the behavior of PEV of the wheel random regular network ($\mathcal{G}_1$), we consider random edge rewiring in $\mathcal{G}_1$ (Fig. \ref{schematic_random_edge_rewiring}) as follows. We remove an edge $e_p \in E$ ($p=1,2,\ldots,|E|$) uniformly at random from $\mathcal{G}_{i}$ and at the same time, introduce an edge uniformly at random in $\mathcal{G}_{i}$ from $e_q^c \in E^c$ ($q=1,2,\ldots,|E^c|$) and record the IPR values (Fig. \ref{schematic_random_edge_rewiring}). 
The new network and the corresponding adjacency matrix are denoted as $\mathcal{G}_{i+1}$ and ${\bf A}_{i+1}$, respectively. Starting from a $\mathcal{G}_{1}$ (constructed using the analytical method), the random edge rewiring process yields a sequence of networks $\{\mathcal{G}_{1},\mathcal{G}_{2},\ldots, \mathcal{G}_{i},\mathcal{G}_{i+1},\ldots,\mathcal{G}_{\tau}\}$ and the corresponding adjacency matrices as $\{{\bf A}_{1},{\bf A}_{2},\ldots, {\bf A}_{i}, {\bf A}_{i+1},\ldots,{\bf A}_{\tau}\}$ where $\tau$ is the total number of edge rewiring performs. Notably, during an edge rewiring, there is a possibility that the network becomes disconnected. However, we allow only those edge rewirings which yield a connected network. Note that the random edge rewiring and the optimized edge rewiring  \cite{evec_localization_2017} are different processes.

\begin{figure}[t]
\begin{center}
\includegraphics[width=3in, height=1.8in]{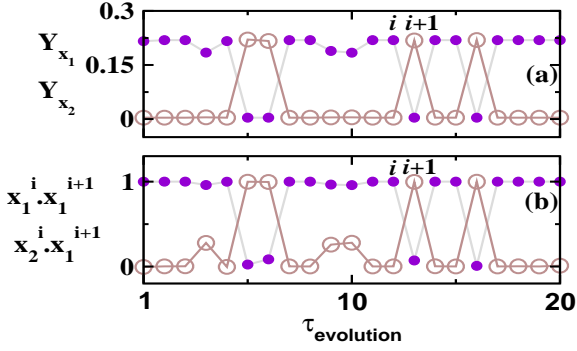}
\caption{(a) Flipping behavior of IPR values of the largest (\textcolor{violet}{$\bullet$}) and second largest (\textcolor{brown}{$\circ$}) eigenvectors. (b) Detection of eigenvalue crossing through dot products $(\bm{ x}_{1}^{i})^{T}\bm{ x}_{1}^{i+1}$ (\textcolor{violet}{$\bullet$}) and $(\bm{ x}_{2}^{i})^{T}\bm{ x}_{1}^{i+1}$ (\textcolor{brown}{$\circ$}) during the random edge rewiring process in wheel random regular network structure. Here, $n=500$ and $\langle k \rangle=10$.}
\label{eigenvalues_IPRs}
\end{center}
\end{figure}
\begin{figure}[t]
\begin{center}
\includegraphics[width=2.8in, height=2.2in]{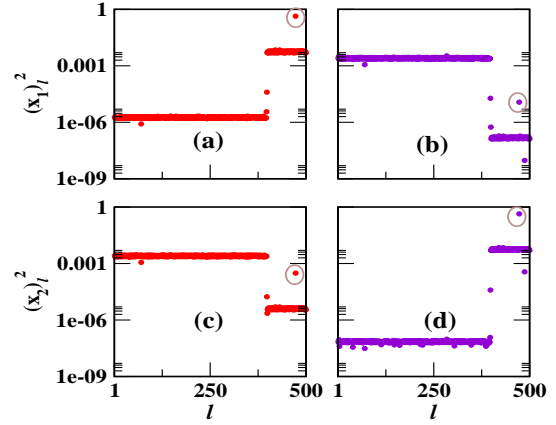}
\caption{Largest two eigenvectors entries in (a) \& (c) wheel random regular network where the entry corresponding to the hub node is marked with a circle, (b) \& (d) after a single edge rewiring of the wheel random regular network. Network parameters are same as in Fig. \ref{eigenvalues_IPRs}.} 
\label{evec_entries}
\end{center}
\end{figure}

 
The random edge rewiring process reveals that removal of an edge connected to the hub node in $\mathcal{W}$ component followed by the addition of that edge to the $\mathcal{R}$ component leads to delocalization of PEV as well as high localization of the second-largest eigenvector ($\bm{x}_2$). This is referred to as sensitivity in PEV (Fig. \ref{eigenvalues_IPRs} (a)). 
In other words, there exists abrupt changes in the $Y_{\bm{x}_{1}}$ and $Y_{\bm{x}_{2}}$ values due to a single edge rewiring. To elaborate this aspect of the abrupt changes in the IPR values as a consequence of a single edge rewiring, we focus on two consecutive networks, say, ${\bf A}_{i}$ and ${\bf A}_{i+1}$ such that ${\bf A}_{i+1}$ is achieved after the single edge rewiring process on ${\bf A}_{i}$. We observe that $\bm{x}_1^{i+1}$ reaches to a delocalized state from a highly localized state at the $(i+1)^{th}$ time step (Fig. \ref{eigenvalues_IPRs} (a)). This abrupt change in the IPR value of $\bm{x}_1^{i+1}$ is accompanied with a high localization of $\bm{x}_2^{i+1}$ from a delocalized state ($\bm{x}_2^{i}$) (Fig. \ref{eigenvalues_IPRs} (a)). Scrutinizing the entries of the eigenvectors corresponding to the largest and the second largest eigenvalues in these two consecutive steps, we find that there exist radical changes in the eigenvector entries (Fig. \ref{evec_entries}). One can observe that, $\bm{ x}_{1}^{i}$ is highly localized with maximum entry corresponding to the hub node (marked with a circle in Fig. \ref{evec_entries}(a)). However, after a single edge rewiring on ${\bf A}_{i}$, ${\bf A}_{i+1}$ has the same structure (except a single edge rewired), $\bm{x}_{1}^{i+1}$ becomes delocalized (Fig. \ref{eigenvalues_IPRs} (a)). The entry corresponding to the hub node for this delocalized $\bm{x}_{1}^{i+1}$ takes a very small value (Fig. \ref{evec_entries}(b)). Notably, for $\bm{x}_{2}^{i+1}$, the entry corresponding to the hub node takes the same value as that of the $\bm{x}_{1}^{i}$ (Fig. \ref{evec_entries}(a) and (d)). This clear flip in the entries of the largest two eigenvectors ($\bm{x}_{1}^{i+1}$ and $\bm{x}_{2}^{i+1}$) affect IPR value of both of them.

Further, an examination of relative positions of the two largest eigenvectors provides insight into the sensitive behavior of the PEV in wheel random regular network structure. To trace the relative position of the largest two eigenvectors in the vector space, we analyze the angle by computing the dot product of two vectors, i.e.,
$$(\bm{ x}_{1}^{i})^{T}\bm{ x}_{1}^{i+1}\;\; \mbox{and}\;\; (\bm{x}_{2}^{i})^{T}\bm{ x}_{1}^{i+1} \;\;\mbox{for}\;i=1,2,\ldots,\tau$$
during the random edge rewiring process. One can see that presence of the flips in IPR values (Fig. \ref{eigenvalues_IPRs}(a)) are reflected in the similar abrupt changes in the dot product values (Fig. \ref{eigenvalues_IPRs}(b)). These abrupt changes in $(\bm{ x}_{1}^{i})^{T}\bm{ x}_{1}^{i+1}$ and $(\bm{ x}_{2}^{i})^{T}\bm{ x}_{1}^{i+1}$ manifest a signature of the eigenvalue crossing phenomenon. The rewiring of an edge connected to the hub node leads to rotation of $\bm{ x}_1$ and $\bm{ x}_2$ by approx. $90^{o}$ (Fig. \ref{eigenvalues_IPRs}(b)). It has already been reported that abrupt changes in the eigenvector entries carry information of the eigenvalue crossing \cite{eigenlevel_crossing_2015, eigenlevel_crossing_2006}. Moreover, it has also been noted that just after the crossing, the eigenvectors become orthogonal to the eigenvectors before the crossing (i.e., $\bm{ x}_{1}^{i} \perp \bm{ x}_{1}^{i+1}$ and $\bm{ x}_{2}^{i} \perp \bm{ x}_{2}^{i+1}$). The largest two eigenvectors satisfy these two criteria mentioned above during the flipping of the IPR values. Further, to confirm the eigenvalue crossing phenomenon, we perform the following experiments. We separate two graph components ($\mathcal{C}_1^{i}$ and $\mathcal{C}_2^{i}$) of $\mathcal{G}_{i}$ corresponding to ${\bf A}_i$ by breaking the existing connection between them, and record the largest two eigenvalues. We observe that the largest two eigenvalues of the $\mathcal{G}_{i}$ remain almost the same as of the largest eigenvalue of the two subgraph components separately 
$$\lambda_1^{\mathcal{C}_1^{i}} \approx  \lambda_1^{\mathcal{G}_{i}},\;\; \lambda_1^{\mathcal{C}_2^{i}} \approx \lambda_2^{\mathcal{G}_{i}} $$ 
Further, one can also notice that
\begin{equation}\label{leading_eig_1}
\lambda_1^{\mathcal{C}_1^{i}} > \lambda_1^{\mathcal{C}_2^{i}} 
\end{equation}
In an another experiment, if we remove an edge from $\mathcal{G}_i$ connected to the hub node in $\mathcal{C}_1^{i}$, and add it between a randomly selected pair of the nodes in  $\mathcal{C}_2^{i}$, this leads to an abrupt change in the localization behavior of PEV. The modified network is denoted as $\mathcal{G}_{i+1}$. This reshuffling of an edge makes $\bm{ x}_{1}^{i+1}$ delocalized and $\bm{ x}_{2}^{i+1}$ 
highly localized (Fig. \ref{eigenvalues_IPRs} (a)). Next, upon separating two components of $\mathcal{G}_{i+1}$, we observe that 
$$ \lambda_1^{\mathcal{C}_2^{i+1}} \approx \lambda_1^{\mathcal{G}_{i+1}} ,\; \lambda_1^{\mathcal{C}_1^{i+1}} \approx \lambda_2^{\mathcal{G}_{i+1}} $$ 
The transition between the localized and the delocalized states for $\bm{ x}_1^{i+1}$ and $\bm{ x}_2^{i+1}$, respectively in $\mathcal{G}_{i+1}$ is accompanied with a change in the $\lambda_1^{\mathcal{C}_{1}^{i+1}}$ value leading to 
\begin{equation}\label{leading_eig_2}
\lambda_1^{\mathcal{C}_{1}^{i+1}} < \lambda_1^{\mathcal{C}_{2}^{i+1}} 
\end{equation}
For both the experiments, the largest eigenvalues of $\mathcal{G}_{i}$ and $\mathcal{G}_{i+1}$ are always greater than the corresponding second largest eigenvalues i.e.,
$$\lambda_1^{\mathcal{G}_{i}} > \lambda_2^{\mathcal{G}_{i}} \;\mbox{and}\; \lambda_1^{\mathcal{G}_{i+1}} > \lambda_2^{\mathcal{G}_{i+1}}$$
which also satisfy the Perron-Frobenius theorem \cite{matrix_analysis}. However, the changes in the largest eigenvalue of the individual components in $\mathcal{G}_{i}$ and $\mathcal{G}_{i+1}$ (Eqs. (\ref{leading_eig_1}) and (\ref{leading_eig_2})) occur due to the eigenvalue crossing. In the other words, for the case of the highly localized PEV of wheel random regular network, the component containing the hub node has the prime contribution in the largest eigenvalue. Note that the analysis in \cite{evec_centrality_loc_2020} (Appendix B) reflects that by obeying $\lambda_1^{\mathcal{W}}>\lambda_1^{\mathcal{R}}$, PEV entries of $\mathcal{G}_{new}$  corresponding to the wheel subgraph contribute more to IPR as compared to those of the random regular graph part.

\subsection{Localization behavior on RNA dynamical model}\label{RNA}
In the previous sections, we investigated the eigenvalue crossing phenomenon and its relation with the sensitive behavior of the PEV of the adjacency matrices. In this section, we turn our attention to show the impact of the eigenvalue crossing phenomenon, caused by single edge rewiring, on the steady-state behavior of a linear dynamical system. We consider the RNA neutral network population linear dynamical model \cite{rna_model_2009,rna_model_2018, rna_neutral_evolution_1999} 
which represents a set of genotypes mapping to the same phenotype form of a neutral network. Nodes in this neutral network correspond to genotypes (sequences), and two nodes are said to be connected if the corresponding sequences differ by a single point mutation. Each node $i$ holds a number $x_i(t)$ from the sequence at time $t$. At each time step, each sequence replicates at a rate $f>1$ and each daughter sequence mutates to one of the $3L$ nearest neighbors with a probability $\mu$, whereas with the probability  $1-\mu$ it does not mutate. Here, $L$ is the sequence length and $0< \mu < 1$. The equations illustrating the dynamics of the population on a network can be given by
\begin{equation}\label{rna_model}
x_i(t+1) = f(1-\mu)x_i(t)+\frac{f\mu}{3L}\sum_{i=1}^{n} a_{ij}x_j(t)
\end{equation}
In the matrix form 
\begin{eqnarray}
\bm{x}(t+1) &= f(1-\mu){\bf I} \bm{x}(t)+\frac{f\mu}{3L}{\bf A}\bm{x}(t) \nonumber\\
&=\biggl[f(1-\mu){\bf I}+\frac{f\mu}{3L}{\bf A}\biggr]\bm{x}(t) \nonumber \\
&={\bf M}\bm{x}(t) \label{rna_model}
\end{eqnarray} 
where ${\bf M}$, ${\bf I}$, and ${\bf A}$ are the transition, identity, and adjacency matrix, respectively. 
\begin{figure}[t]
\begin{center}
\includegraphics[width=3in, height=1.8in]{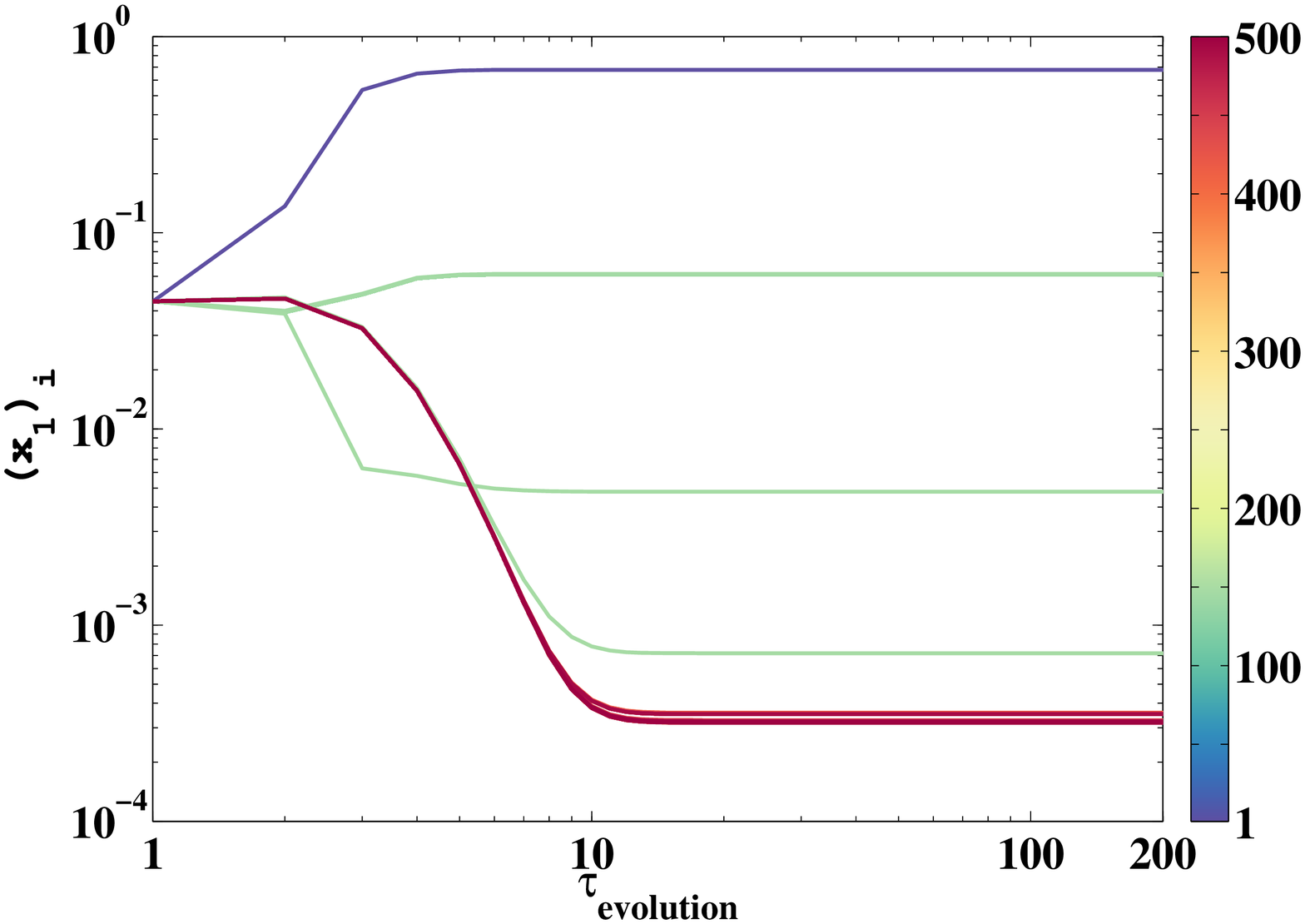}
\caption{Evolution of the steady-state vector of the RNA neutral network model. Starting with a uniform state vector, we perform the power iteration method to reach the steady-state vector. Due to the localized PEV, the hub node contributes more to the dynamical process, and the rest of them have a tiny contribution. Here, $n=500$, $\mu=0.5$, $f=2.6$, $L=18$. We perform the power iteration method for $300000$ iterations and store the PEV after each $1500$ steps.}
\label{RNA_loc}
\end{center}
\end{figure}
For the above model, the steady-state vector is obtained from the PEV of the transition matrix \cite{rna_model_2009}. Importantly, all the eigenvectors of ${\bf A}$ and ${\bf M}$ are same which can easily be shown from Eq. (\ref{rna_model}) as follows 
\begin{eqnarray}
{\bf M}\bm{x}_i^{\bf A} &= f(1-\mu){\bf I}\bm{x}_i^{\bf A}+\frac{f\mu}{3L}{\bf A}\bm{x}_i^{\bf A} \nonumber\\
 &= f(1-\mu)\bm{x}_i^{\bf A} + \frac{f\mu}{3L}\lambda_i^{\bf A}\bm{x}_i^{\bf A} \nonumber \\
 &= \lambda_i^{\bf M}\bm{x}_i^{\bf A} \label{rna_model_pev}
\end{eqnarray}
where $\lambda_i^{\bf M} = f(1-\mu)+\frac{f\mu}{3L}\lambda_i^{\bf A}$, $\lambda_i^{\bf M}$ and $\lambda_i^{\bf A}$ denotes the eigenvalues and $\bm{x}_i^{\bf M}$ and $\bm{x}_i^{\bf A}$ are the eigenvectors of ${\bf M}$ and ${\bf A}$ respectively. Further, $\lambda_1^{\bf M}$ is the asymptotic growth rate of the population and from Eq. (\ref{rna_model_pev}) one can observe that limit distribution of population or the steady-state vector of the transition matrix is solely determined by the PEV of the adjacency matrix \cite{rna_neutral_evolution_1999,pevec_nat_phys_2013}.

\begin{figure}[t]
\begin{center}
\includegraphics[width=3in, height=1.9in]{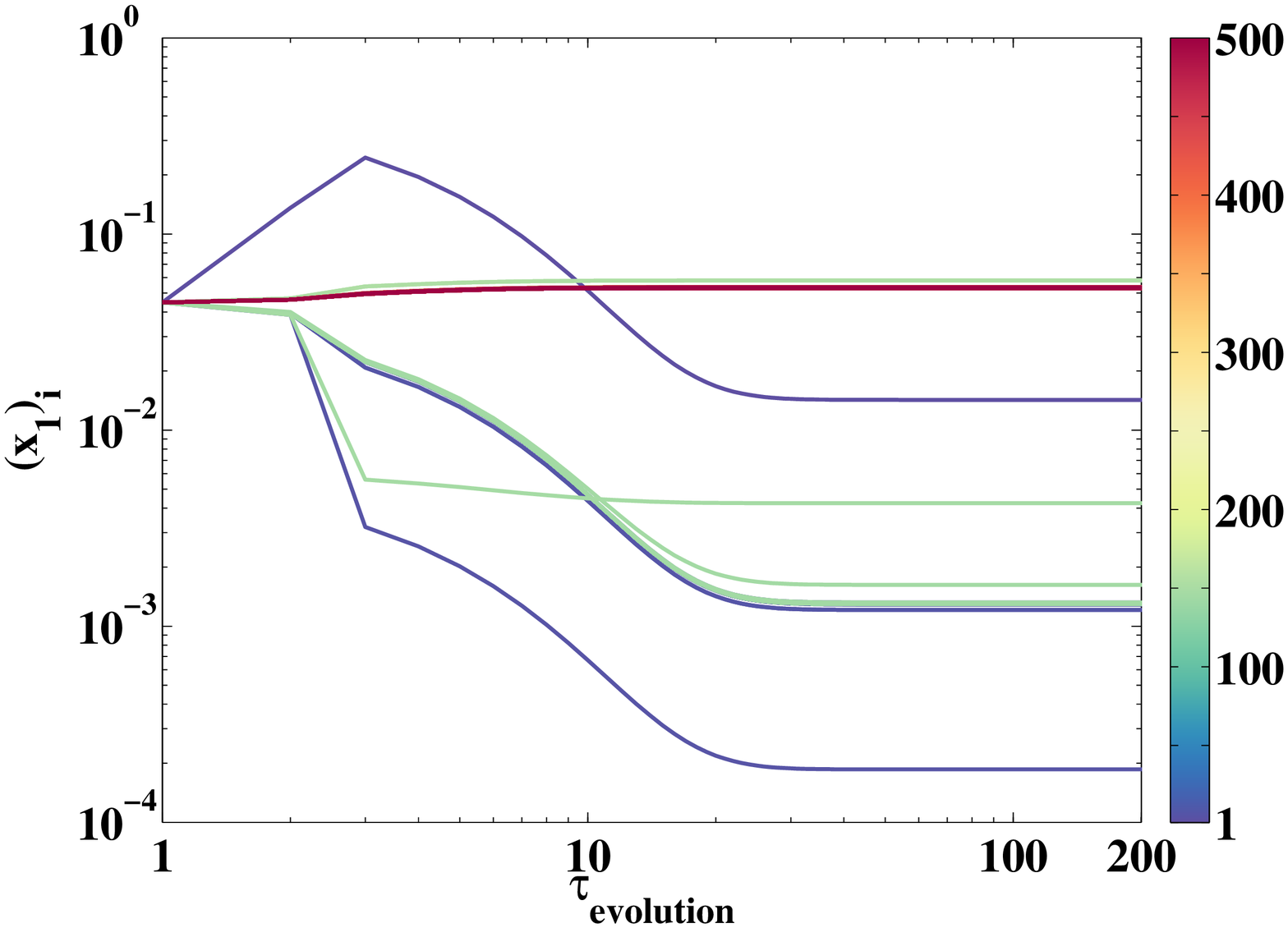}
\caption{Evolution of the steady-state vector of the RNA neutral network model on the wheel-random structure where we rewire an edge connected to the hub node. Starting from a uniform state vector, we perform the power iteration method to reach the steady-state vector \cite{colormap_2015}. Due to the delocalized PEV, there exists a drastic change in the steady-state of the dynamical process. Model parameters are the same as in Fig. \ref{RNA_loc}.}
\label{RNA_dloc}
\end{center}
\end{figure}
We perform the power iteration method on ${\bf M}$ with an initial population distribution vector having all the entries same. Considering ${\bf A}$ as the adjacency matrix corresponding to the wheel random regular structure with $\lambda_{1}^{\mathcal{W}} > \lambda_1^{\mathcal{R}}$, 
maximum contribution to the dynamical process comes from a single node (Fig. \ref{RNA_loc}). In the wheel random regular network, we rewire an edge connected to the hub node and add it to the random regular structure, and the new transition matrix is denoted by ${\bf M}^{'}$. We again perform the power iteration method on ${\bf M}^{'}$ with the initial population distribution vector, which has all the entries same. 

One can observe drastic changes in the steady-state vector of the RNA model arising due to the eigenvalue crossing phenomenon (Figs. \ref{RNA_loc} and \ref{RNA_dloc}). The two largest eigenvalues of the network remain close to each other; however, there exist changes in the individual eigenvalue relation leading to the change in the behavior of the steady-state. To avoid this sensitive dependence of the steady-state arising due to a single edge rewiring, we increase the largest eigenvalue of the wheel graph component by either increasing the size of the wheel graph or by increasing the average degree of the regular graph component as learned from the analytical approach discussed in subsection \ref{Analytical}. 
The prime aim of including the RNA neutral network population dynamical model is to demonstrate the peculiar behavior of the steady-state on the wheel random regular network from the mathematical perspective. However, existing literature on the neutral network population dynamical model show signatures that the evolution of the RNA secondary structure population over the neutral network is not random but tends to concentrate at highly connected parts of the network \cite{rna_model_2009,rna_neutral_evolution_1999,modeling_evolutionary_landscape_1999}.
On the other hand, in the RNA neutral networks, nodes are genotype, which are connected by an edge if there is a single point mutation. Next, a single edge rewiring can be perceived as a change in the mutation point of two sequences. Few previous papers \cite{rna_model_2009,rna_neutral_evolution_1999} point out that neutrality is essential in the evolution of quasi-species and there is a chance of random mutation. However, in reality, whether there can exist a drastic change in the steady-state of RNA population dynamical model due to single edge rewiring (mutation) needs a more in-depth biological study, which can be one of the directions of the future investigation. 

Although the wheel random regular structure is quite special, which may be difficult to observe for real-world systems. However, we know that many real-world networks follow power-law degree distributions and thus contain a set of very large degree nodes that naturally form imperfect wheel graphs (i.e., star, friendship). Our study offers a platform to have a better understanding of the behavior of linear dynamical processes of real-world systems in the steady-state as well as to relate them with the structural properties of underlying network structures. Note that the dynamical system used here is a simplified and discrete-time version of the Eigen's molecular-evolution model \cite{rna_neutral_evolution_1999}. 

\section{Conclusion}\label{conclusion}
This article provides an analytical method for the construction of a highly localized network structure for a given set of network parameters. In other words, by mapping the eigenvalue relationship between the individual subgraph components to a cubic equation and solving it analytically, we find the subgraph component size for the construction of PEV localized networks. We show that a highly localized network structure is accompanied with sensitivity in the localization behavior of PEV against a single edge rewiring. Moreover, we find evidences for eigenvalue crossing phenomena as a consequence of the single edge rewiring, thereby providing an origin to the sensitive behavior of PEV localization. Finally, we substantiate the eigenvalue crossing phenomenon observed in localized networks by using the RNA neutral network population linear-dynamical model. Additionally, the eigenvalue crossing phenomenon can be extended for the maximal entropy random walk model \cite{maximal_entropy_random_walk_2009} which is a recreated version of the quantum walk with many applications such image analysis, tampering detection \cite{tampering_detection}, object localization \cite{object_localization}, tractography problem \cite{tractography_problem}, or detecting visual saliency regions \cite{visual_saliency_regions}. 

Although the structure of the wheel random regular network is far from those of the real-world networks, few unique properties (localized PEV, the existence of sensitivity, presence of a hub node with its size related to the largest eigenvalues of the individual component) of the localized networks can act as benchmarks for further applications and future theoretical analysis. Note that instead of using a wheel graph, we can also use a star graph to construct $\mathcal{G}_{new}$ for the construction of the  PEV localized networks.

The analysis of the localization behavior of PEV is useful to understand steady-state behaviors of many linear-dynamical processes ranging from epidemic spreading to biochemical dynamics \cite{Goltsev_prl2012, pevec_nat_phys_2013}. A linear-dynamical system represented by a network having localized PEV indicates that a few nodes have significant contributions to that dynamical process, and the rest of the other nodes have very tiny contributions. Similarly, for a delocalized PEV, all the nodes have almost the same amount of contribution to the underlying linear-dynamical process. Consequently, the network properties which enhance the PEV localization can implicitly restrict the linear-dynamics to a smaller section of the network. An understanding of the network properties having highly localized PEV is therefore essential to engineering the system's architecture to restrict or to spread of the dynamics. Further, the PEV localization can be useful in understanding the eigenvector centrality measures in networks \cite{evec_centrality_loc_2020}. Here, we have focused only on the adjacency matrices with binary entries, which are different from the matrices used in the Anderson localization and several other matrix representations of networks (e.g., Laplacian, Jacobian, Hessian matrices) \cite{loc_laplacian_matrix, laplacian_random_matrix_2008,network_dynamics_loc_lap_2018,jacobian_matrix_loc_2014}. It will be interesting to use the framework developed here to analyze other matrix representations of complex networks. 

All the data and codes used in this paper are available at GitHub repository \cite{codes_data_eigval_crossing}.


%



\ifCLASSOPTIONcompsoc
  \section*{Acknowledgment}
\fi
SJ acknowledges CSIR, Govt. of India (25(0293)/18/EMR-II), and DAE, Govt. of India (37(3)/14/11/2018-BRNS/37131) grants for financial support. PP acknowledges CSIR, Govt. of India grant (09/1022(0070)/2019-EMR-I) for SRF fellowship. We are thankful to Manavendra Mahato (Indian Institute of Technology Indore) for useful discussions on the eigenvalue crossing phenomenon. PP is indebted to Beresford N. Parlett (University of California, Berkeley) for the helpful suggestion of testing the gap between close  eigenvalues through the spectrum slicing method. PP thanks to all Complex Systems Lab members at IIT Indore for useful discussions and support.

\ifCLASSOPTIONcaptionsoff
  \newpage
\fi

\end{document}


%
\title{Appendix: From Spectra to Localized Networks: A Reverse Engineering Approach}

\author{Priodyuti~Pradhan 
        and~Sarika~Jalan
\IEEEcompsocitemizethanks{\IEEEcompsocthanksitem Priodyuti~Pradhan was with the Complex Systems Lab, Discipline of 
Physics, Indian Institute of Technology Indore, Khandwa Road, Simrol, Indore-453552, India. 
E-mail: priodyutipradhan@gmail.com
\IEEEcompsocthanksitem Sarika Jalan is with the Complex Systems Lab, Discipline of 
Physics, Indian Institute of Technology Indore, Khandwa Road, Simrol, Indore-453552, India.
E-mail: sarikajalan9@gmail.com
}
}

\maketitle

\IEEEdisplaynontitleabstractindextext

\IEEEpeerreviewmaketitle

\newcommand{\beginsupplement}{%
\setcounter{equation}{0}
        \renewcommand{\theequation}{A\arabic{equation}}
        \setcounter{table}{0}
        \renewcommand{\thetable}{A\arabic{table}}%
        \setcounter{figure}{0}
        \renewcommand{\thefigure}{A\arabic{figure}}%
     }

\section*{Discriminant analysis}
\beginsupplement

The section analyzes the discriminant of Eq. (12) 
and provides the bounds for the wheel graph size ($n_1$) as a function of $n$. To achieve, we first find the range of $m$ values and their relations with the behavior of discriminant ($\Delta$). Then, we calculate the bounds for the roots and calculate the bounds for $n_1$. We rewrite the discriminant in Eq. (12) of main article 
\begin{equation}
\begin{split}
\Delta &=\frac{\beta^2}{4}+\frac{\alpha^3}{27}\\ 
 &= (m-n\sigma-p)^2 -\biggl(\frac{n}{3}+q\biggr)^3 \label{discriment_analysis_apx} 
\end{split}
\end{equation}
where $\sigma=(1-\frac{\epsilon}{3})$, $p = \frac{\epsilon^{3}+9\epsilon^{2}+36\epsilon}{27}$, and $q=\frac{\epsilon^2+6\epsilon+6}{9}$. We consider connected network and choose $m$ in between $n+1$ to $n(n-1)/2$ where $n \geq 49$. 

\noindent {\bf Case (i) [$\Delta=0$]:}
To find out the value of $m$ for which $\Delta=0$, we solve,
\begin{equation}\label{discriment_analysis_eq_c1_apx}
(m-n\sigma-p)^2 -\biggl(\frac{n}{3}+q\biggr)^3 =0 
\end{equation}
Solving the quadratic equation for $m$, we get two roots $m = (n\sigma+p) \pm \frac{(n+3q)\sqrt{3(n+3q)}}{9}$ for which $\Delta = 0$. However, we know that $m$ should always be a positive quantity, hence we choose the positive root.
\begin{equation}\label{delta_cases1_apx}
m = (n\sigma+p) + \frac{(n+3q)\sqrt{3(n+3q)}}{9} 
\end{equation}
Moreover, for a given $n$, $m$ is always be a positive integer but from Eq. (\ref{delta_cases1_apx}), $m$ is a real value with fractional part. Hence, $\Delta=0$ can never appear for our case.

\noindent {\bf Case (ii) [$\Delta>0$]:} Now, as $m$ should be a positive integer we add $1$ to Eq. (\ref{delta_cases1_apx}) and get the lower bound for $m$ value as follows
\begin{equation}\label{delta_cases2_apx}\small
\biggl \lceil (n\sigma+p+1) + \frac{(n+3q)\sqrt{3(n+3q)}}{9} \biggr \rceil \leq m \leq \frac{n(n-1)}{2} 
\end{equation}
for which $\Delta > 0$. Now, we substitute Eq. (\ref{discriment_analysis_apx}) in Eq. (12), 
and we have
{\small \begin{equation}\nonumber 
\begin{split}
\Delta_1 &= \Biggl[-(m-n\sigma-p)+\sqrt{(m-n\sigma-p)^2-\biggl(\frac{n}{3}+q\biggr)^3}\Biggr]^{1/3}\\
\Delta_2 &= \Biggl[-(m-n\sigma-p)-\sqrt{(m-n\sigma-p)^2-\biggl(\frac{n}{3}+q\biggr)^3}\Biggr]^{1/3} 
\end{split}
\end{equation}}\normalsize
Further, for the range of $m$ values mentioned in Eq. (\ref{delta_cases2_apx}), $\Delta>0$ or $(m-n\sigma-p)^2-(\frac{n}{3}+q)^3>0$, thus  $(m-n\sigma-p)>\sqrt{(m-n\sigma-p)^2-(\frac{n}{3}+q)^3}$. Therefore, $\frac{\sqrt{(m-n\sigma-p)^2-(\frac{n}{3}+q)^3}}{m-n\sigma-p}<1$ and hence using bionomial approximation we get
\begin{eqnarray}
\Delta_1 &\approx -(m-n\sigma-p)^{1/3}\Biggl[1-\frac{\sqrt{(m-n\sigma-p)^2-(\frac{n}{3}+q)^3}}{3(m-n\sigma-p)}\Biggr] \nonumber\\
\Delta_2 &\approx -(m-n\sigma-p)^{1/3}\Biggl[1+\frac{\sqrt{(m-n\sigma-p)^2-(\frac{n}{3}+q)^3}}{3(m-n\sigma-p)}\Biggr] \nonumber
\end{eqnarray}
Therefore, from Eq. (11) 
and using the above two relations we get,
\begin{equation}\label{root_delta_greater}
\kappa_1 =-2(m-n\sigma-p)^{1/3}+\frac{6-2\epsilon}{3}\\ 
\end{equation}
Further, from Eq. (\ref{root_delta_greater}) with the help of inequality in Eq. (\ref{delta_cases2_apx}), we get lower bound for $\kappa_1$ using the bionomial approximation as follows
\begin{eqnarray}
\kappa_1 &> -2\biggl(\frac{n(n-1)}{2}-n\sigma-p \biggr)^\frac{1}{3}+\frac{6-2\epsilon}{3} \nonumber\\
      &= -2\biggl(\frac{n^2}{2}-\frac{n(9-2\epsilon)}{6}-p\biggr)^\frac{1}{3} + \frac{6-2\epsilon}{3} \nonumber\\
     & \;\mbox{for}\; 0<\epsilon \ll 1 \nonumber \\
     & \approx -2^{2/3}n^{2/3}\biggl(1-\frac{1}{n}\biggr)+2 \nonumber\\
     & \;\mbox{for}\; n\rightarrow \infty  \nonumber\\
        &\approx -(2n)^{2/3}+2 \nonumber
\end{eqnarray}
Similarly, we calculate the upper bound for $\kappa_1$ from Eqs. (\ref{delta_cases2_apx}) and (\ref{root_delta_greater}) as follows
$$\kappa_1 < -\frac{2}{\sqrt{3}}\sqrt{n}+2$$ 
Hence, combining the above two cases for $\Delta>0$ we have bounds for $\kappa_1$ as follows 
\begin{equation}
-  (2n)^{2/3}+2< \kappa_1 < -\frac{2}{\sqrt{3}}\sqrt{n}+2 
\end{equation}
and finally from Eq. (7), 
we get bounds for $n_1$ as follows
\begin{equation}
\frac{4}{3}n-\frac{4}{\sqrt{3}}\sqrt{n} < n_1^{\kappa_1}< (2n)^{4/3}-4n^{2/3}
\end{equation}
From the above, we conclude that for a given $n$ value as $m$ varies in the range given in Eq. (\ref{delta_cases2_apx}), size of the wheel graph varies in the above range. Finally, we show that $\frac{4}{3}n-\frac{4}{\sqrt{3}}\sqrt{n}>n$ for $n\geq 49$ and $(2n)^{4/3}-4n^{2/3}>n$ for $n \geq 4$. Hence, for $n\geq 49$, size of the wheel graph exceeds the given $n$. Thus, we can not use $\kappa_1$ for the wheel graph size calculation from Eq. (7). 

\noindent {\bf Case (iii) [$\Delta<0$]:} 
Subtracting $1$ from Eq. (\ref{delta_cases1_apx}), we get upper bound for $m$ 
\begin{equation}\label{delta_cases1}\small
n+1 \leq m \leq \biggl \lceil (n\sigma+p-1)+ \frac{(n+3q)\sqrt{3(n+3q)}}{9}\biggr \rceil
\end{equation}
for which $\Delta<0$. Now from Eq. (12) we get
{\scriptsize \begin{equation}
\begin{split}
\Delta_1 &=  \sqrt[3]{-\beta/2+ \sqrt{\Delta}}\\
&=\Biggl[-(m-n\sigma-p)+\sqrt{(m-n\sigma-p)^2-\biggl(\frac{n}{3}+q\biggr)^3}\Biggr]^{1/3}
\end{split}
\end{equation}}\normalsize
For the above range of $m$ in Eq. (\ref{delta_cases1}), $\Delta<0$ or $(m-n\sigma-p)^2-\biggl(\frac{n}{3}+q\biggr)^3<0$. Hence, we have
{\scriptsize \begin{equation}
\begin{split}
\Delta_1 &=\Biggl[-(m-n\sigma-p)+\sqrt{-\biggl(\biggl(\frac{n}{3}+q\biggr)^3-(m-n\sigma-p)^2\biggr)}\Biggr]^{1/3}\\
&=\Biggl[-(m-n\sigma-p)+i\sqrt{\biggl(\frac{n}{3}+q\biggr)^3-(m-n\sigma-p)^2}\Biggr]^{1/3}
\end{split}
\end{equation}}\normalsize
where $i=\sqrt{-1}$ and $\biggl(\frac{n}{3}+q\biggr)^3-(m-n\sigma-p)^2>0$. Therefore, following the inequality in Eq. (\ref{delta_cases1}), from Eq. (12)  
we get
\begin{eqnarray}
\Delta_1 =z_1^{1/3}\qquad \mbox{and}\qquad \Delta_2 =z_2^{1/3} \nonumber  
\end{eqnarray}
where
{\scriptsize \begin{eqnarray}
z_1 &= \Biggl[-(m-n\sigma-p)+i\sqrt{\biggl(\frac{n}{3}+q\biggr)^3-(m-n\sigma-p)^2}\Biggr]\nonumber \\
z_2 &= \biggl[-(m-n\sigma-p)-i\sqrt{\biggl(\frac{n}{3}+q\biggr)^3-(m-n\sigma-p)^2}\biggr] \label{delta_case3}
\end{eqnarray}}\normalsize
Hence, $\Delta_1$ and $\Delta_2$ are the cubic roots of complex numbers $z_1$ and $z_2$ respectively. Therefore,
in the polar form
\begin{eqnarray}
z_1 &= r_{z_1}[\cos \theta_{z_1} + i \sin \theta_{z_1}] \nonumber\\
z_2 &= r_{z_2}[\cos \theta_{z_2} + i \sin \theta_{z_2}] \nonumber
\end{eqnarray}
and the cubic roots of $z_1$ and $z_2$ can be calculated as
\begin{eqnarray}
\Delta_{1}^{s} &= \sqrt[3]{r_{z_1}}\Biggl[\cos \frac{2\pi s+\theta_{z_1}}{3} + i\sin \frac{2\pi s+\theta_{z_1}}{3} \Biggr],\; s=0,1,2 \nonumber\\
\Delta_{2}^{s} &= \sqrt[3]{r_{z_2}}\Biggl[\cos \frac{2\pi s+\theta_{z_2}}{3} + i\sin \frac{2\pi s+\theta_{z_2}}{3} \Biggr],\; s=0,1,2 \nonumber
\end{eqnarray}
and hence from Eq. (11) we get
\begin{equation}
\begin{split}
\kappa_1 &= \Delta_1^{s} + \Delta_2^{s} - \frac{b}{3} \\
&=\sqrt[3]{r_{z_1}}\Biggl[\cos \frac{2\pi s+\theta_{z_1}}{3} + i\sin \frac{2\pi s+\theta_{z_1}}{3} \Biggr]+ \\
&\sqrt[3]{r_{z_2}}\Biggl[\cos \frac{2\pi s+\theta_{z_2}}{3} + i\sin \frac{2\pi s+\theta_{z_2}}{3} \Biggr]- \frac{b}{3} \label{k1_1}
\end{split}
\end{equation}
To simplify the above equation, we perform the following steps. From Eq. (\ref{delta_case3}), we calculate
{\scriptsize \begin{equation}\nonumber
\begin{split}
r_{z_1} &= \sqrt{(-(m-n\sigma-p))^2+\biggl(\sqrt{\biggl(\frac{n}{3}+q\biggr)^3-(m-n\sigma-p)^2}\biggr)^2} \\
&= \biggl(\frac{n}{3}+q\biggr)^\frac{3}{2} 
\end{split}
\end{equation}}\normalsize
Similarly, from Eq. (\ref{delta_case3}) we also get, $r_{z_2} = \biggl(\frac{n}{3}+q\biggr)^\frac{3}{2}$. Hence,
\begin{equation}\label{modulus}
r_{z_1} =r_{z_2}= \biggl(\frac{n}{3}+q\biggr)^\frac{3}{2}
\end{equation}
Now, one can see that for the range of $m$ value in Eq. (\ref{delta_cases1}), $(m-n\sigma-p) >0 $ and $\sqrt{(\frac{n}{3}+q)^3-(m-n\sigma-p)^2} > 0$ for $0<\epsilon \ll 1$. Hence, $z_1$ and $z_2$ in Eq. (\ref{delta_case3}) belongs to the second and third quadrant of the Argand plane and complex conjugate to each other. We find the principal value for the argument in the range of $(-\pi, \pi]$ \cite{complex_variable}. Hence, the argument becomes 
\begin{equation}\label{theta_relation}
\theta_{z_2} =-\theta_{z_1}
\end{equation}
Now, from Eq. (\ref{k1_1}) by using the relations in Eqs. (\ref{modulus}) and (\ref{theta_relation}) we get
\begin{equation}\label{k1_2}
\kappa_1 =2\sqrt[3]{r_{z_1}} \cos \frac{\theta_{z_1}}{3} \Biggl[ \cos \frac{2\pi s}{3} +i \sin \frac{2\pi s}{3}\Biggr]- \frac{b}{3}\\
\end{equation}
Further, it is known $\Delta<0$ provides three unequal real roots, hence, $\kappa_1$ should be a real value \cite{cubic_roots}. One can see that we get a real value for $s=0$ and complex number for other $s$ values. Finally, for $s=0$, from Eq. (\ref{k1_2}) 
we get 
\begin{equation}\label{delta_less_1}
\kappa_1 =2\biggl(\frac{n}{3}+q\biggr)^\frac{1}{2} \cos \frac{\theta_{z_1}}{3}+\frac{6-2\epsilon}{3}
\end{equation}
and similarly from Eq. (11) 
by using the relation in Eqs. (\ref{modulus}), (\ref{theta_relation}) and for $s=0$, we get
\begin{figure}[t]
\begin{center}
\includegraphics[width=3.1in, height=1.3in]{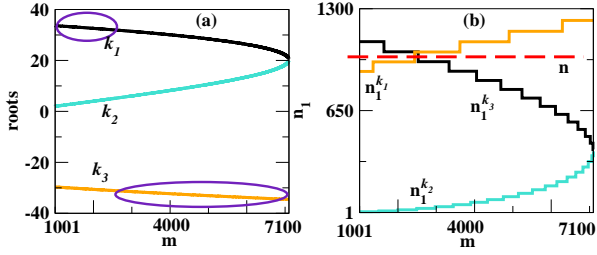}
\caption{In the cubic equation for the coefficient $n=1000$, $\epsilon=0.00002$ and for different values of $m$ in the range given by Eq. (\ref{delta_cases1}). (a) One can observe nature of three unequal real roots for $\Delta<0$. (b) Behavior of the wheel graph component size calculated from Eq. (\ref{vertices_wheel_relation}) and for three different roots denoted as $n_1^{\kappa_1}$, $n_1^{\kappa_2}$ and $n_1^{\kappa_3}$ respectively. We can obseve that for sparse network $n_1^{\kappa_1}$ is larger than $n$. On the other hand as network becomes dense, $n_1^{\kappa_3}$ becomes larger than $n$. $n_1^{\kappa_2}$ is always leser than $n$.}
\label{nature_of_roots}
\end{center}
\end{figure}
\begin{eqnarray}
\kappa_2 = 2\biggl(\frac{n}{3}+q\biggr)^\frac{1}{2} \sin \biggl(\frac{\theta_{z_1}}{3} - \frac{\pi}{6}\biggr)+\frac{6-2\epsilon}{3} \nonumber \\
\kappa_3 = -2\biggl(\frac{n}{3}+q\biggr)^\frac{1}{2} \sin \biggl(\frac{\theta_{z_1}}{3} + \frac{\pi}{6}\biggr)+\frac{6-2\epsilon}{3} \label{delta_less_2}
\end{eqnarray}
Next, we calculate the lower and upper bounds for the roots in the range of $m$ for a given $n$ in Eq. (\ref{delta_cases1}).
We know that $z_1$ is in second quadrant, thus,  $\frac{\pi}{2} <\theta_{z_1}<\pi$, implies $\frac{\pi}{6} <\frac{\theta_{z_1}}{3}<\frac{\pi}{3}$, hence, $\frac{1}{2}<\cos \frac{\theta_{z_1}}{3}<\frac{\sqrt{3}}{2}$ and which is positive. Further, $0 <\frac{\theta_{z_1}}{3}-\frac{\pi}{6}<\frac{\pi}{6}$ implies that $0< \sin (\frac{\theta_{z_1}}{3} - \frac{\pi}{6}) <\frac{1}{2}$. Finally, $ \frac{\pi}{3} <\frac{\theta_{z_1}}{3}+\frac{\pi}{6}<\frac{\pi}{2}$ implies that $\frac{\sqrt{3}}{2}< \sin (\frac{\theta_{z_1}}{3} + \frac{\pi}{6}) <1$. Further, we find the lower and upper bound for the roots from Eqs. (\ref{delta_less_1}) and (\ref{delta_less_2}) using the binomial approximation for $0<\epsilon\ll 1$ and $n\rightarrow \infty$ as follows
\begin{eqnarray}
\frac{1}{\sqrt{3}}\sqrt{n}+2 < \mathfrak{\kappa}_1 < \sqrt{n}+2 \nonumber\\
2 <\mathfrak{\kappa}_2 <\frac{1}{\sqrt{3}}\sqrt{n}+2 \nonumber \\
 -\frac{2}{\sqrt{3}}\sqrt{n} + 2 < \mathfrak{\kappa}_3 < -\sqrt{n} + 2 \nonumber 
\end{eqnarray}
Finally, use the lower and upper bounds of $\kappa_i$ and calculate the bounds of $n_1$ in Eq. (7) 
as follows
\begin{eqnarray}
\frac{1}{3}n+\frac{2}{\sqrt{3}}\sqrt{n} < n_1^{\kappa_1} < n+2\sqrt{n} \nonumber\\
1 < n_1^{\kappa_2} < \frac{1}{3}n+\frac{2}{\sqrt{3}}\sqrt{n} \nonumber \\
n -2\sqrt{n}< n_1^{\kappa_3} < \frac{4}{3}n-\frac{4}{\sqrt{3}}\sqrt{n} \nonumber
\end{eqnarray}
From the above $\frac{1}{3}n+\frac{2}{\sqrt{3}}\sqrt{n}>n$ for $n<3$, $n+2\sqrt{n}>n$ for $n>0$, and finally $\frac{4}{3}n-\frac{4}{\sqrt{3}}\sqrt{n}>n$, $n > 48$. Hence, if we choose $n \geq 49$, $n_1^{\kappa_2}$ will always be less than $n$.

We numerically vary $m$ in the range in Eq. (\ref{delta_cases1}) and examine the behavior of three different roots (Fig. \ref{nature_of_roots}(a)) and their corresponding $n_1$ values (Fig. \ref{nature_of_roots}(b)). One can observe that for a small region, size of $n_1^{\kappa_1}$ and $n_1^{\kappa_3}$ exceeds the given $n$ (depicted by a horizontal dotted line in Fig. \ref{nature_of_roots}(b)). Importantly, the bounds obtained from the analysis are in good agreement with the numerical results and indicate that for sparse networks small portion of the $\kappa_1$ cannot be used to find wheel graph size (Fig. \ref{nature_of_roots}(a) marked with an ellipse). Consequently, for dense networks, $\kappa_3$ can not be used for the wheel graph size calculation (Fig. \ref{nature_of_roots}(a) marked with an ellipse) and $\kappa_2$ always works well. Hence, we use $\kappa_2$ as well as $\kappa_1$ or $\kappa_3$ to calculate the wheel and random regular component size to construct $\mathcal{G}_{new}$.

Note that for $\kappa = 2$ and $0<\epsilon \ll 1$, we get $n_1< 4$ and which is not a valid size for constructing a $\mathcal{W}$ component. The smallest value to construct a $\mathcal{W}$ component is $n_1= 4$ and to satisfy Eq. (6), we need $\kappa = 3$ for $0<\epsilon \ll 1$. On the other hand, for a random regular network having $n_2$ nodes, the maximum possible degree is $n_2-1$. However, to perform the random edge rewiring process, we should keep the maximum degree to be $n_2-2$ and thus $3 \leq \kappa \leq n_2-2$.

\section{Eigenvalue crossing}
Eigenvalue crossing or sometimes say energy level crossing is a phenomenon where two eigenvalues (energy) crosses to each other due to some operations in the system. Exactly why does the crossing happen, which is not exactly clear, and it is system dependent and varies from system to system \cite{eigenlevel_crossing_2006}. Here, we find that a network structure having highly localized PEV can show eigenvalue crossing phenomena due to the rewiring of edges inside the network structure. Here, rewiring is the operation in the network structure (system). As we know, to detect eigenvalues crossing, the researcher checks the position of the corresponding eigenvector before and after the operations performed. It has been reported that before and after the crossing, the corresponding eigenvector is orthogonal \cite{eigenlevel_crossing_2015}. To check the position, we consider the dot product value. Note that as we consider connected network, hence, $\lambda_1>\lambda_2$ (non-degenerate) from Perron-Frobenius theorem and numerically verified using spectrum slicing method (\cite{symmetric_eigenvalue_book}, chapter 3).

\ifCLASSOPTIONcompsoc